\documentclass[a4paper,onecolumn,12pt,assc]{IEEEtranASSC}

\ifCLASSINFOpdf 
 \usepackage[pdftex]{graphicx}
\else
 \usepackage[dvips]{graphicx}
\fi

\usepackage{threeparttable}
\newcommand{\kms}{\,kms$^{-1}$}

\usepackage{url}

\begin{document}

\title{Planets in Spin-Orbit Misalignment and the Search for Stellar Companions}

\author{\IEEEauthorblockN{Brett C. Addison\IEEEauthorrefmark{1}\IEEEauthorrefmark{2},
C. G. Tinney\IEEEauthorrefmark{1}\IEEEauthorrefmark{2},
Duncan J. Wright\IEEEauthorrefmark{1}\IEEEauthorrefmark{2},
Graeme Salter\IEEEauthorrefmark{1}\IEEEauthorrefmark{2},
Daniel Bayliss\IEEEauthorrefmark{3} and
George Zhou\IEEEauthorrefmark{3}}

\IEEEauthorblockA{\IEEEauthorrefmark{1}
School of Physics, University of New South Wales, Sydney, NSW 2052, Australia; b.addison@unsw.edu.au}

\IEEEauthorblockA{\IEEEauthorrefmark{2}
Australian Centre for Astrobiology, University of New South Wales, NSW 2052, Australia}

\IEEEauthorblockA{\IEEEauthorrefmark{3}
Research School of Astronomy and Astrophysics, Australian National University, Canberra, ACT 2611, Australia}}

\maketitle

\vspace{-0.4cm}
\begin{abstract}
The discovery of giant planets orbiting close to their host stars was one of the most unexpected results of early exoplanetary science. Astronomers have since found that a significant fraction of these `Hot Jupiters' move on orbits substantially misaligned with the rotation axis of their host star. We recently reported the measurement of the spin-orbit misalignment for WASP-79b by using data from the 3.9~m Anglo-Australian Telescope. Contemporary models of planetary formation produce planets on nearly coplanar orbits with respect to their host star's equator. We discuss the mechanisms which could drive planets into spin-orbit misalignment. The most commonly proposed being the Kozai mechanism, which requires the presence of a distant, massive companion to the star-planet system. We therefore describe a volume-limited direct-imaging survey of Hot Jupiter systems with measured spin-orbit angles, to search for the presence of stellar companions and test the Kozai hypothesis.
\end{abstract}

\begin{IEEEkeywords}
planets and satellites: dynamical evolution and stability – stars: individual (WASP-79) – techniques: radial velocities and direct imaging
\end{IEEEkeywords}

\vspace{-0.5cm}
\section*{Introduction}

Exoplanetary science is possibly the most exciting and rapidly developing branch of modern astronomy. Just over 1000 planets\footnote{\url{http://exoplanet.eu}, as of 2014 January 15. There is some debate on the fraction of planets that truly exist. The other main online exoplanet database, \url{http://exoplanets.org/}, list the total confirmed planets at just 763 as of 2014 January 15.} have been discovered to date, mainly through radial velocity and transit searches. A detailed discussion of the various methods used to detect exoplanets is beyond the scope of this paper, but for more information, we direct the interested reader to \cite{2011exha.book.....P}. In addition to finding new planets, a detailed analysis of their structure, composition, and other bulk properties is needed to understand the processes involved in their formation and migration. It is possible to probe these processes by measuring the sky-projected spin-orbit alignment (or obliquity) of exoplanetary systems. This is the angle between the planetary orbital plane and the spin vector of the host star. This is done through spectroscopic measurements of the Rossiter-McLaughlin effect (first measured for eclipsing binaries (\cite{1924ApJ....60...15R},\cite{1924ApJ....60...22M}) and since extended to exoplanets \cite{2000A&A...359L..13Q} including the recently measured planet WASP-79b \cite{2013ApJ...774L...9A}). The observed effect is caused by the modification of the stellar spectrum as a transiting planet occults a small region of the stellar disk of its host star, causing a radial velocity anomaly, due to asymmetric distortions in the rotationally broadened stellar line profiles \cite{2005ApJ...622.1118O}. 

The radial velocity anomaly is sketched in Figure 1. For prograde orbits, a planet will first transit across the portion of its host star's disk that is rotating towards the observer (i.e. blue-shifted), blocking light from that hemisphere. This results in the observer receiving a greater fraction of the total flux of the star from the hemisphere that is rotating away (red-shifted) from the observer than that rotating towards the observer. The blue-shifted portion of the rotationally broadened stellar lines will appear to have less absorption, resulting in the line profile centroids being red-shifted and a positive radial velocity anomaly. A negative velocity anomaly will occur during the second half of the transit as the planet moves across the hemisphere rotating away from the observer. For retrograde orbits, a planet will transit across the red-shifted hemisphere first, resulting in the inverse velocity anomaly. By measuring the shape and magnitude of the Rossiter-McLaughlin effect, it is possible to determine the inclination of a planet's orbit relative to the spin axis of its host star. 

Giant planets are thought to form within the proto-planetary disk that surrounds a protostar through the core-accretion process \cite{1996Icar..124...62P}. This model predicts that Jovian planets should form several AU\footnote{An astronomical unit, or AU, is a standard unit for measuring distances in astronomy. 1~AU is slightly less than 150 million kilometers, and is approximately the mean distance between the Earth and the Sun.} away from the protostar where the proto-planetary disk is sufficiently cool for icy volatiles to exist and to slowly accrete into planetesimals. The planetesimals continue to accrete material, growing until they reach a critical mass of $\sim5 - 10$ $M_{\oplus}$, at which point they rapidly accrete gas from the surrounding disk. This leads to the formation of a large gaseous envelope of hydrogen and helium. Accretion halts and planet formation comes to an end once the gas in the local disk is exhausted or blown away by the protostar as it reaches the main sequence phase (\cite{1996Icar..124...62P},\cite{2005A&A...434..343A}).

There are two complicating factors to this simple model of planet formation. First, gas giant planets are found well inside 1~AU where they cannot have formed in situ, implying that they must have migrated in from their birthplaces (i.e. disk migration \cite{2010MNRAS.401.1505B}). Secondly, many Hot Jupiters at small orbital radii are observed to be in spin-orbit misalignment \cite{2012ApJ...757...18A}. This is unexpected as the proto-planetary disk from which planets form should be well aligned with the plane of the protostar's equator (e.g., \cite{2005ApJ...622.1118O}; \cite{2005ApJ...631.1215W}), suggesting that either the migration process, or post-migration evolution of the planet's orbit, has driven planets into spin-orbit misalignment. Several mechanisms have been proposed to produce these anomalies, such as Kozai resonances \cite{2011Natur.473..187N}, secular chaos \cite{2011ApJ...735..109W}, and planet-planet scattering \cite{2008ApJ...686..580C}. 

To date, of the 74 planetary systems\footnote{\label{RM_foot}This study has made use of Ren\'{e} Heller's Holt-Rossiter-McLaughlin Encyclopaedia which was last updated on 2013 November; \url{http://www.physics.mcmaster.ca/~rheller/index.html}} with measured obliquities, 33 show substantial misalignments ($>22.5^{\circ}$), 10 of which are in nearly polar orbits, and 7 are in retrograde orbits. With such a large fraction of planets in spin-orbit misalignment, there is a clear need to understand the physical mechanisms that are producing these systems.

\begin{figure}[t]
\centering
\includegraphics[width=0.66\linewidth]{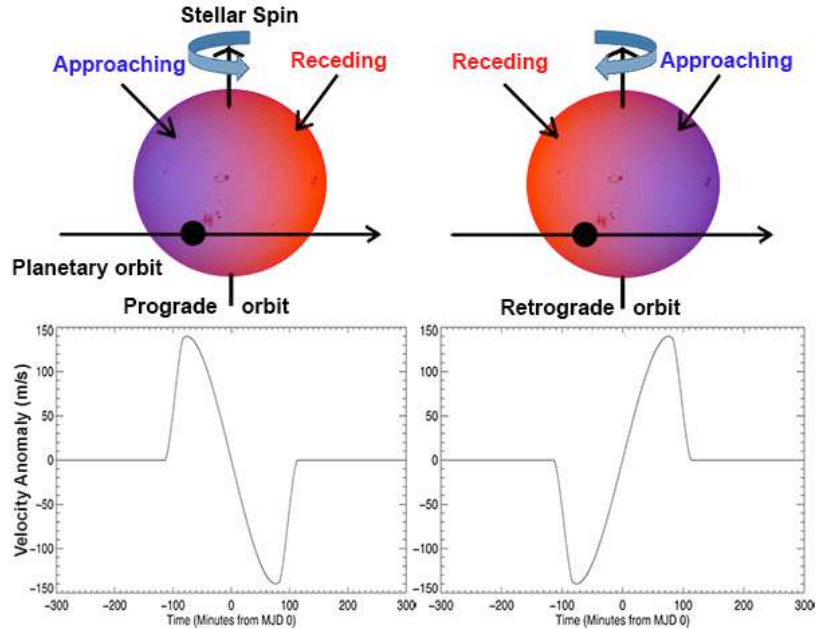}
\caption[LoF entry]{Top: Artist's impression of a transiting planet on a prograde and a retrograde orbit (for clarity, the motion of the planet across its host star is the same for both prograde and retrograde orbits while the rotation direction of the star is reversed for the retrograde orbit). The stellar hemisphere shaded blue represents the side rotating towards the observer with its light blue-shifted. The stellar hemisphere shaded red represents the side rotating away from the observer with its light red-shifted.

Bottom: The modeled radial velocities for the two orbits showing the differing Rossiter-McLaughlin effect as a planet transits across the disk of its host star.}
\label{Figure_prograde_retrograde}
\end{figure}

\section*{Spin-Orbit Misalignment of WASP-79b}
WASP-79b is a bloated Hot Jupiter that was recently discovered through the Wide Angle Search for Planets (WASP) \cite{2012A&A...547A..61S}. We determined the spin-orbit misalignment of WASP-79b through spectroscopic measurements of the Rossiter-McLaughlin effect, using high-precision radial velocity observations taken during the transit on the night of 2012 December 23, using the CYCLOPS2\footnote{\url{http://www.phys.unsw.edu.au/~cgt/CYCLOPS/CYCLOPS_2.html}} optical-fiber bundle system feeding the UCLES eschelle spectrograph on the Anglo-Australian Telescope at Siding Spring Observatory. Raw spectra were reduced using custom routines developed by the authors and were wavelength calibrated from a thorium-argon observation taken at the beginning of the night and from thorium-xenon spectra taken using the simultaneous calibration fiber during each object exposure \cite{2013ApJ...774L...9A}. We used the IRAF\footnote{IRAF is distributed by the National Optical Astronomy Observatories, which are operated by the Association of Universities for Research in Astronomy, Inc., under cooperative agreement with the National Science Foundation.} task, \emph{fxcor}, to compute radial velocities by cross-correlation with a spectrum of a bright template star (HD86264) of similar spectral type. Details on the reduction and data analysis of WASP-79 can be found in \cite{2013ApJ...774L...9A}.

We developed the Exoplanetary Orbital Simulation and Analysis Model (ExOSAM) to fit our radial velocities of WASP-79b with a model of the Rossiter-McLaughlin effect using the Hirano et al. analytical approach \cite{2010ApJ...709..458H}. The best-fitting values for the projected spin-orbit angle, $\lambda$, and the projected stellar rotational velocity, $v\sin i_{\star}$\footnote{$v$ is the absolute rotational velocity of the star and $i_{\star}$ is the inclination of the star's rotational axis to the observers line of sight}, along with the uncertainties in these parameters were derived using a grid search and minimizing $\chi^{2}$ between the observed radial velocities and modeled radial velocities \cite{2013ApJ...774L...9A}. 

Two solutions for the stellar parameters of WASP-79 have been derived from photometric data \cite{2012A&A...547A..61S} -- one with WASP-79 on the main sequence ($R_{\star}=1.64 \pm 0.08$ $R_{\odot}$) and one with it evolved just off the main sequence ($R_{\star}=1.91 \pm 0.09$ $R_{\odot}$). This results in two sets of solutions for the various other system parameters including $\lambda$ and $v\sin i_{\star}$ which are given in Table 1. Our results for the projected spin-orbit alignment and stellar rotation velocity, using the main sequence parameters, are $\lambda = -106^{+19}_{-13}$\,$^{\circ}$ and $v\sin i_{\star} = 17.5^{+3.1}_{-3.0}$\kms. For the non-main sequence case, $\lambda = -84^{+23}_{-30}$\,$^{\circ}$ and $v\sin i_{\star} = 16.0^{+3.7}_{-3.7}$\kms. In both cases, WASP-79b is in a significantly misaligned orbit.

The main sequence solution appears to be the most likely one \cite{2012A&A...547A..61S}, as the main sequence lifetime of a star is significantly longer than its post-main sequence lifetime \cite{1992A&AS...96..269S}. Since photometric data neither prefers the main sequence or the evolved solution for WASP-79, and given that a star is far more likely to be observed on the main sequence, we focus on the main sequence solution. Figure 2 shows the velocity anomaly for the main sequence solution with the observed velocities over-plotted on the left. A positive hump-shaped anomaly due to the Rossiter-McLaughlin effect is clearly apparent in our velocities. This implies that the planet must be in a nearly polar orbit and transits across only the blue-shifted hemisphere (or the side rotating toward us) as depicted in the illustration on the right side of figure 2. 

\begin{table*}[t]
\begin{threeparttable}[t]
\caption{The relevant system parameters of WASP-79 as a main sequence and evolved star. The full listing of system parameters for WASP-79 can be found in \cite{2013ApJ...774L...9A}. Here, ms denotes the model that assumes WASP-79 to be a main sequence star, while non-ms denotes the model that assumes it is instead an evolved star.}
\centering
\begin{tabular}{l c c}
\hline\hline \\ [-2.0ex]
Parameter & Value (ms) & Value (non-ms) \\ [0.5ex]
\hline\hline \\ [-2.0ex]
{\textit{Parameters as given by Smalley et al. (2012)}} \\
\hline \\ [-2.0ex]
Mid-transit epoch (2400000-HJD), $T_{0}$ & $56285.03589 \pm 0.00200$ & $56285.03739 \pm 0.00300$ \\
Orbital period, $P$ & $3.6623817 \pm 0.0000050$ d & $3.6623866 \pm 0.0000085$ d \\
Semi-major axis, $a$ & $0.0539 \pm 0.0009$ AU & $0.0535 \pm 0.0008$ AU \\
Orbital inclination, $i$ & $85.4 \pm 0.6^{\circ}$ & $83.3 \pm 0.5^{\circ}$ \\
Impact parameter, $b$ & $0.570 \pm 0.052$ & $0.706 \pm 0.031$ \\
Transit depth, $(R_{P}/R_{\star})^{2}$ & $0.01148 \pm 0.00051$ & $0.01268 \pm 0.00063$ \\
Orbital eccentricity, $e$ & 0.0 (assumed) & 0.0 (assumed) \\
Stellar mass, $M_{\star}$ & $1.56 \pm 0.09$ $M_{\odot}$ & $1.52 \pm 0.07$ $M_{\odot}$ \\
Stellar radius, $R_{\star}$ & $1.64 \pm 0.08$ $R_{\odot}$ & $1.91 \pm 0.09$ $R_{\odot}$ \\
Planet mass, $M_{P}$ & $0.90 \pm 0.09$ $M_{J}$ & $0.90 \pm 0.08$ $M_{J}$ \\
Planet radius, $R_{P}$ & $1.70 \pm 0.11$ $R_{J}$ & $2.09 \pm 0.14$ $R_{J}$ \\ [0.5ex]
\hline \\ [-2.0ex]
\textit{Parameters determined from model fit using our velocities} \\ [0.5ex]
\hline \\ [-2.0ex]
Projected obliquity angle, $\lambda$ & $-106^{+19}_{-13}$ $^{\circ}$ & $-84^{+23}_{-30}$ $^{\circ}$ \\ [0.5ex]
Projected stellar rotation velocity, $v\sin i_{\star}$ & $17.5^{+3.1}_{-3.0}$ \kms\ & $16.0^{+3.7}_{-3.7}$ \kms\ \\ [0.5ex]
\hline 
\end{tabular}
\end{threeparttable}%
\label{table:parameters}
\end{table*}

\begin{figure}[t]
\centering
\includegraphics[width=0.9\linewidth]{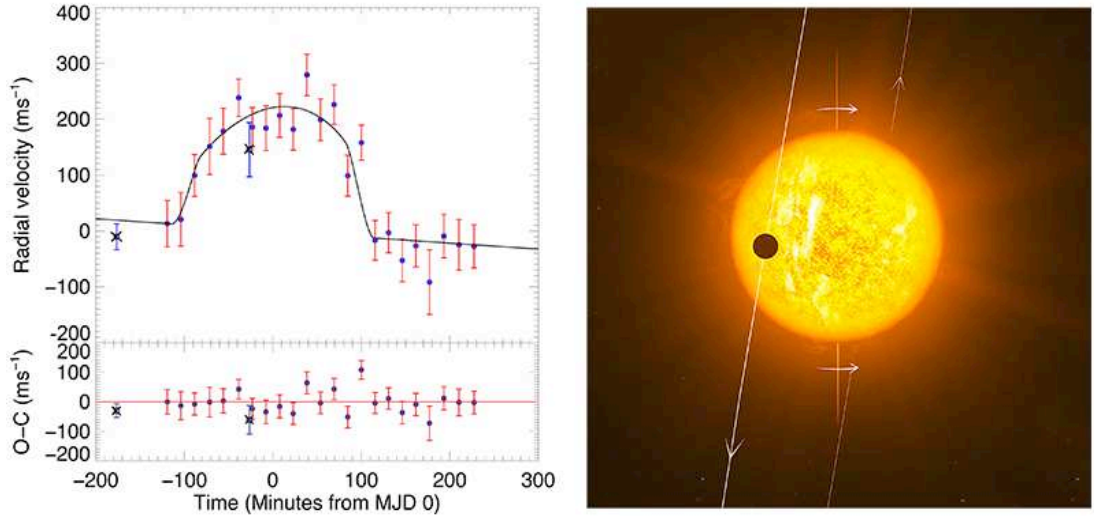}
\caption[LoF entry]{Left: Spectroscopic radial velocities of the WASP-79b transit taken on 2012 December 23 using the CYCLOPS2 fiber-bundle on the Anglo-Australian Telescope \cite{2013ApJ...774L...9A}. Velocities from just before, during, and after the transit are plotted as a function of time along with the best fitting model (for the main sequence parameters) and corresponding residuals. The filled blue circles with red error bars are velocities we measured with our estimated uncertainty. The two black circles with an x and with blue error bars are previously published velocities using the quoted uncertainties \cite{2012A&A...547A..61S}.

Right: Artist impression of the WASP-79b polar orbit. Image credit: Brett Addison (modified version of the original artist impression of the WASP-8b retrograde orbit by ESO/L. Cal\c{c}ada).}
\label{Figure_1_WASP-79b_transit}
\end{figure}

\section*{Observed Trends of Systems in Spin-Orbit Misalignment}
One early trend observed for planetary systems with measured obliquities is that planets in spin-orbit misalignment tend to orbit hot stars with $T_{eff} \geq 6250$~K while planets in spin-orbit alignment tend to orbit cooler stars with $T_{eff} < 6250$~K \cite{2010ApJ...718L.145W}. The correlation between spin-orbit misalignments and stellar effective temperature may be related to the thickness of the stellar convective zone, since it has been proposed that the convective zone could act to dampen orbital obliquities over time through enhancing planet-star tidal interactions \cite{2010ApJ...718L.145W}. Hot stars have a thin convective layer and are expected to have very long obliquity dampening timescales, typically orders of magnitude longer than the main sequence lifetime of the star. On the other hand, stars cooler than $6250$~K have much shorter obliquity dampening timescales that are fractions of the main sequence lifetime. 

This interpretation of the dichotomy observed for aligned and misaligned systems has been supported by a new study \cite{2012ApJ...757...18A}, which measured the spin-orbit alignments for 14 new systems and computed the obliquity dampening timescales for these systems and for 39 previously published systems. A positive correlation was found between obliquity and stellar temperature as well as a positive correlation between obliquity and the timescales for dampening obliquity. The reasoning for this observed correlation is similar to that of an earlier study \cite{2010ApJ...718L.145W}.

Recent studies of multi-planet transiting systems have revealed low stellar obliquities for five systems \cite{2013ApJ...771...11A} and a high obliquity in one system \cite{2013Sci...342..331H}. The study on low stellar obliquity systems that were known at the time of \cite{2013ApJ...771...11A} publication (shortly before the high obliquity system was announced \cite{2013Sci...342..331H}) suggested that the migration mechanism(s) responsible for producing Hot Jupiters is fundamentally different from the mechanism(s) producing compact close-in multi-planet systems \cite{2013ApJ...771...11A}. The \cite{2013ApJ...771...11A} study proposes that multi-planet systems likely migrated due to disk-planet interactions while Hot Jupiters experienced dynamical perturbations during migration from Kozai resonances or planet-planet scattering. 

This view has been challenged by the discovery of a significant spin-orbit misalignment (true obliquity angle $\psi > 37^{\circ}$) for the Kepler-56 multi-planet system \cite{2013Sci...342..331H}. This result illustrates that spin-orbit misalignments are not restricted to Hot Jupiter systems. The presence of an additional massive body (planet, brown-dwarf, or low mass star) in a wide orbit has been detected from radial velocity measurements which reveal a long term systematic trend \cite{2013Sci...342..331H}. The high stellar obliquity and the presence of a third companion is interpreted as evidence for a scenario in which torques from the outer massive companion drive the inner planets into co-planer orbits that are misaligned with the spin-axis of the host star \cite{2013Sci...342..331H}. It is unlikely that the high obliquity of Kepler-56 is due to the Kozai resonances or planet-planet scattering. Whether a similar scenario to the one proposed for Kepler-56 is responsible for spin-orbit misalignments seen in Hot Jupiters is yet to be determined. Therefore, measuring the obliquities of multi-planet systems is key to determining if such misalignments are common or rare, and whether the mechanism(s) driving misalignments are similar to or different from Hot Jupiters.

We calculated the tidal dissipation timescale for WASP-79b \cite{2013ApJ...774L...9A} to test its consistency with the overall trends that have been observed for other systems. Using either of the two methods presented in \cite{2012ApJ...757...18A}, we found that the tidal dissipation timescale for WASP-79b is between $\tau_{mcz}=1.6\times10^{11}$~yr to $\tau_{RA}=3.3\times10^{15}$~yr, which is longer than that calculated for $80\%$ of the systems examined in \cite{2013ApJ...771...11A}. WASP-79 has an effective temperature of $T_{eff}=6600\pm100$K, which is above the $T_{eff}>6250$K threshold claimed for planetary systems displaying high orbital obliquities and consistent with the very long tidal dissipation timescale we have calculated for this system. Figures 4 and 5 show the projected orbital obliquity of WASP-79b and those planets presented in \cite{2012ApJ...757...18A} as a function of the effective stellar temperature and the relative tidal-dissipation timescale, respectively.

\begin{figure}[t]
\centering
\includegraphics[width=0.75\linewidth]{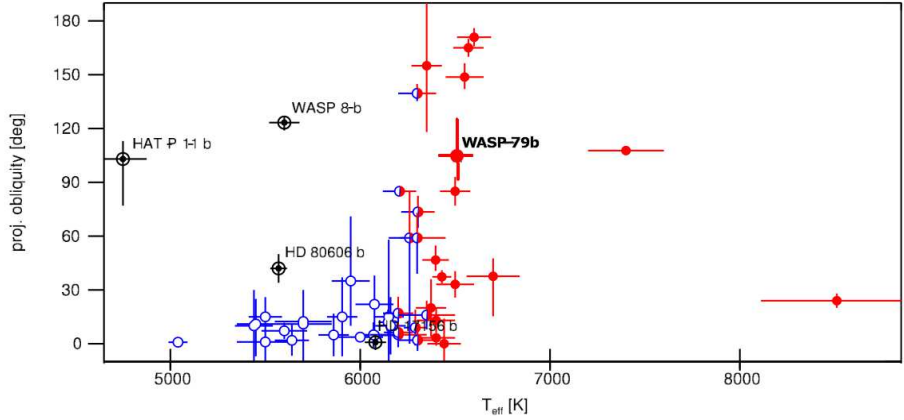}
\caption{Projected orbital obliquity as a function of stellar effective temperature \cite{2012ApJ...757...18A}. The filled red circles with red error bars are for stars that have temperatures higher than $6250$~K. The blue unfilled circles with blue error bars show stars with effective temperatures lower than $6250$~K. The circles that are half red and blue show stars that have measured effective temperatures consistent with $6250$~K from the $1\sigma$ interval. Systems which harbor planets with masses $<0.2M_{J}$ or orbital periods longer than 7 days are shown by a black circle with a black dot in the middle and black error bars. WASP-79b has been included in this figure.}
\label{Figure_3_eff_temp_obliquity}
\end{figure}

\begin{figure}[t]
\centering
\includegraphics[width=0.70\linewidth]{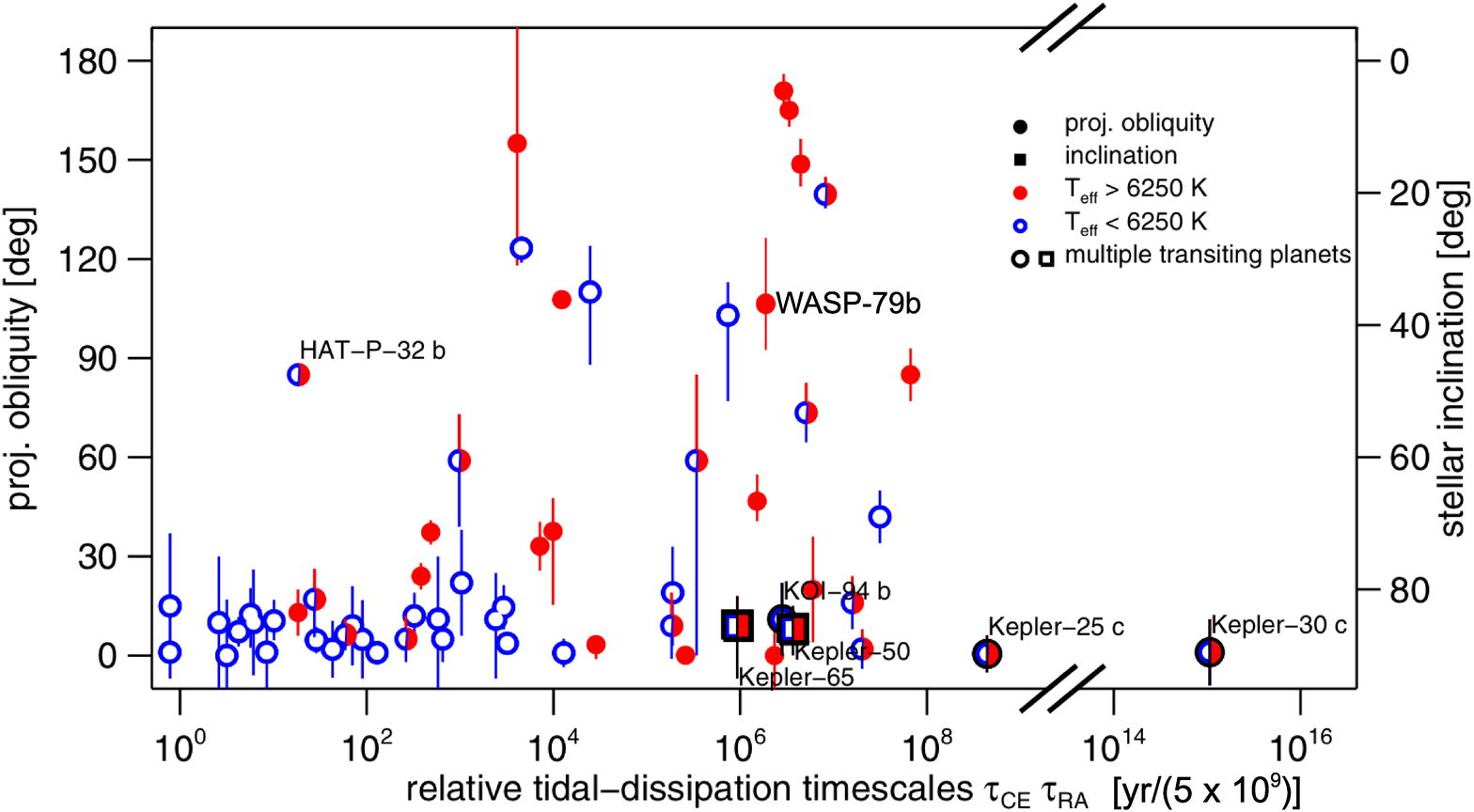}
\caption{Projected orbital obliquity as a function of the relative alignment timescale for stars with either convective (CE) or radiative envelopes (RA) calibrated from binary studies \cite{2013ApJ...771...11A}. The same symbols are used as in Figure 3 with the addition of multiple transiting planets, indicated by the dark black borders. Systems with measured projected obliquity ($\lambda$) are shown as circles while stellar inclinations ($i_{\star}$) are shown as squares. We have calculated the tidal-dissipation timescale for WASP-79b and include it in this figure.}
\label{Figure_4_tidal_dissipation_obliquity}
\end{figure}

\section*{Mechanisms Driving Spin-Orbit Misalignments}
Several mechanisms have been proposed to explain the spin-orbit misalignments observed in many Hot Jupiter systems. These include disk only migration, dynamical mechanisms such planet-planet scattering, secular chaos, Kozai resonances, or some combination of these during or post-disk migration \cite{2008ApJ...678..498N}. Additional scenarios for producing misalignments include stellar internal gravity wave modulation and primordial circumstellar disk misalignments.

Disk migration was one of the earliest and most widely accepted mechanisms invoked to explain the origin of Hot Jupiters \cite{1996Natur.380..606L}. It is understood that gas giant planets form several AU from their host star. However, interactions with the circumstellar disk during and after their formation can induce type 1 and type 2 inward migration \cite{1997Icar..126..261W}. In type 1 migration, planets produce spiral density waves in the circumstellar disk, which results in non-zero tidal torques that drive them inward \cite{1997Icar..126..261W}. Planets more massive than about $10 M_{\bigoplus}$ will quickly clear a gap in the circumstellar disk as they migrate, leading to a transition from type 1 to type 2 migration. Type 2 migration involves the exchange of angular momentum between the planet and the disk as material from the disk enters the gap \cite{1997Icar..126..261W}. If Hot Jupiters are exclusively the result of type 1 and type 2 disk migration, one would expect that their orbits to be well aligned with the stellar spin axis of their host star \cite{2010MNRAS.401.1505B}. This, however, is simply not the case as nearly $45$\% of Hot Jupiters show significant spin-orbit misalignments\footnote{This study has made use of Ren\'{e} Heller's Holt-Rossiter-McLaughlin Encyclopaedia which was last updated on 2013 November; \url{http://www.physics.mcmaster.ca/~rheller/index.html}}.

The standard disk migration model may also play a role in the production of misaligned Hot Jupiters, if the spin-orbit misalignments are the result of the internal gravity wave mechanism altering the orientation of the spin-axis of the host star (\cite{2012ApJ...758L...6R}; \cite{2013ApJ...772...21R}). This model predicts that internal gravity waves are generated at the boundary between the convective core and radiative envelope in relatively hot and massive stars. These waves transport angular momentum outwards towards the surface of the star, causing the surface to rotate at a different speed, and even in a different direction than the interior. Furthermore, the waves themselves can vary over time causing changes in the rotation speed and direction of the stellar surface. While this model does seem to produce systems in spin-orbit misalignment, it has not been determined whether the distribution of obliquities it produces matches the observed population of Hot Jupiters. Other critical tests of this hypothesis include the measurement of time variations in spin-orbit alignments which the internal gravity wave model predicts will occur on timescales of as little as 9 to 1000 rotational periods (\cite{2012ApJ...758L...6R}; \cite{2013ApJ...772...21R}). In addition, measuring the obliquities of multi-planet systems around hot stars with the same misalignments would provide strong evidence for the internal gravity wave model and disfavor dynamical mechanisms. This is because the observed misalignments would be created through changes in the star's spin-axis and would result in all of the planets in a multi-planetary system having nearly co-planar orbits. Misalignments produced through dynamical mechanisms would result in planets on various orbital planes and with different observed obliquities.

An additional scenario in which disk migration will play a key role in producing misaligned Hot Jupiters is through the primordial misalignments of proto-planetary disks \cite{2012Natur.491..418B}. It has been shown that short period misaligned planetary orbits can be the natural product of disk migration in binary systems if the disk is misaligned. If the orbit of a distant stellar companion is inclined by at least $45^{\circ}$ to the plane of the disk, then gravitational torques from the companion will drive the disk into misalignment with the spin-axes of its host stars \cite{2012Natur.491..418B}. Planets forming from such a disk will have misaligned orbits. Finding stellar companions around misaligned systems and determining the distribution of obliquities produced by this model will test the validity of this mechanism.

A non-disk migration mechanism proposed for driving spin-orbit misalignments is planet-planet scattering. If there were two or more planets in a multiple planet system in initially unstable orbits, strong gravitational perturbations between them during close encounters can lead to the ejection of one or more planets and the inward migration of the surviving planets. The planets that migrate inwards will have their orbital eccentricity and obliquity increased until tidally circularized if the periastron distance reaches to within a few stellar radii \cite{2008ApJ...686..580C}. This mechanism is a leading candidate for explaining the occurrence of giant planets in highly eccentric orbits and has had some success in reproducing the incidence and distribution of these planets as well as the architecture of the Solar System \cite{2009ApJ...699L..88R}. Solar System dynamics studies using the NICE model have revealed that Jupiter, Saturn, Uranus, and Neptune have experienced planet-planet scattering episodes in the early history of the Solar System approximately $60$~My to $1.1$~Gy \cite{2008Icar..196..258L}. The NICE model has been able to explain the distribution of the Kuiper belt objects and the outward migration of Uranus and Neptune through planet-planet scattering. It is widely believed that the giant planets of the early Solar System formed on circular and coplanar orbits that were packed significantly closer together (5.5 -- $14$~AU compared to $19$~AU and $30$~AU for present day locations of Uranus and Neptune respectively \cite{2008Icar..196..258L}). The initially stable orbits of the Solar System giant planets became destabilized when Jupiter and Saturn crossed their mutual 1:2 mean-motion resonance. This increased their eccentricity slightly allowing the eccentricities of Uranus and Neptune to increase to the point where they had close encounters with each other and migrated outward to their present day locations \cite{2008Icar..196..258L}. The NICE model thus highlights the importance of planet-planet interactions in shaping the Solar System into the architecture we see today and is likely involved in the diversity of observed exoplanetary orbits. Planet-planet scattering does have some shortcomings. The primary shortcoming of this mechanism is that the scattering process tends to be quite violent and sudden, which impedes the ability of slow processes such as tides to halt the inward migration of planets into their host star \cite{2008ApJ...678..498N}.

Secular chaos is another non-disk migration mechanism where planetary orbits evolve over long timescales (significantly longer than the orbital periods of the planets) due to small gravitational interactions that occur in systems that have three or more well-spaced planets that (generally) are not in strong mean-motion-resonances \cite{2013arXiv1311.1214L} (though secular effects can occur for some objects trapped in mean-motion resonances \cite{2004MNRAS.355..321H}). The inner most planet in such a system will lose angular momentum (but not orbital energy) to an outer planet that will drive its pericenter towards the star as its eccentricity increases while at the same time increase its orbital obliquity \cite{2011ApJ...735..109W}. Tidal circularization will then dampen the eccentricity, creating a Hot Jupiter. One success of this model is that it correctly predicts the 3-day orbital period pile-up observed for Hot Jupiters (\cite{2011ApJ...735..109W}; \cite{2013arXiv1311.1214L}). Despite the model's success in this prediction, it hardly produces planets with obliquities $>90^{\circ}$ and no planets in retrograde orbits \cite{2013arXiv1311.1214L}, counter to the observed population of Hot Jupiter obliquities.

Kozai resonance is the most commonly adopted mechanism to explain planetary systems in spin-orbit misalignment. This mechanism involves the gravitational interaction between a planet and an outer stellar companion that has an orbit that is highly inclined relative to the orbital plane of the planet and is orbiting at large separations (up to several hundred AU though small separations are also possible \cite{1999ssd..book.....M}) from the central star. The gravitational interactions between the two objects induce Kozai oscillations which increases the inclination and decreases the eccentricity for one object while the other object's eccentricity increases and its inclination decreases. Eccentricity and inclination are therefore anti-correlated during Kozai cycles and are described by the Kozai integral \cite{1999ssd..book.....M} $I_{K}$ 
\begin{equation}
I_{K}=\sqrt{1-e^{2}}\cos{i}
\end{equation}
which remains constant through this process. If the eccentricity of the planet increases to the point were its periastron distance is only a few stellar radii from its host star, it will raise tides on its host star's surface. This will cause the semi-major axis to shrink and orbit to circularize through tidal dissipation during periastron passages \cite{2007ApJ...670..820W}. 

It has been found that $\sim 30\%$ of the observed Hot Jupiter population and up to $100\%$ of the misaligned systems could be produced through this mechanism from modeling the dynamical effect of Kozai resonances on Jupiter-like planets from distant stellar companions \cite{2012ApJ...754L..36N}. It is also one of only two migration mechanisms (the other being secular chaos) which can naturally explain the 3-day orbital period pile-up observed for Hot Jupiters \cite{2011ApJ...735..109W}. If indeed the Kozai mechanism were responsible for the spin-orbit misaligned systems, one would expect to find stellar or massive substellar companions to most if not all of them. Additionally, it has been found that the formation of Hot Jupiters through the Kozai mechanism is significantly suppressed for close binaries compare to the efficiency for more distant stellar companions ($>500$~AU) \cite{2012ApJ...754L..36N}. This conclusion is also in agreement with the observational evidence which suggests that Hot Jupiter systems in close binaries ($<100$~AU) are less common \cite{2011IAUS..276..409E}. Therefore, searching for distant stellar companions, as described in the next section, will provide an important test of this mechanism.

\section*{A Test for the Kozai Mechanism: Searching for Stellar Companions}
Direct imaging surveys of systems with measured spin-orbit alignments can provide a critical test of the hypothesis that misalignments are driven by the Kozai mechanism. With current technology, it is feasible to survey a substantial sample of nearby stars ($d \leq 250$~pc) for the presence of stellar and substellar companions at separations as small as $\sim100$~AU. If companions are found preferentially in systems with misaligned Hot Jupiters (as opposed to systems with aligned Hot Jupiters), then the Kozai hypothesis will have substantial support as the dominant driver of these misalignments. 

Companions can be sought using two obvious techniques. One way is through their radial velocity impacts on the host star, similar to radial velocity searches for exoplanets. A caveat of this method is that the timescales required to detect companions orbiting beyond $5$~AU is greater than $10$~yr. This is not very feasible for finding long period companions in a reasonable time frame. Alternatively, the method we are proposing is to search for companions through direct imaging surveys. With the exception of the closest stars, companions at separations of less than $\sim100$~AU will be difficult to detect due to the achievable contrast being insufficient for detecting faint companions. Despite this limitation, it is thought that stellar companions are not likely to reside in orbits less than $100-500$~AU in systems with giant planets (\cite{2012ApJ...754L..36N}; \cite{2011IAUS..276..409E}) as stellar companions can significantly retard planetesimal accretion (\cite{2010ASSL..366...19E}; \cite{2008ASPC..398..179E}). However, other studies have shown that perturbations from moderately close-in massive companions are usually not strong enough to fully halt accretion (\cite{2007arXiv0705.3113M}; \cite{2006ApJ...641.1148B}). Therefore giant planets are expected to be less frequent (but not completely absent) around binaries with separations of less than $100$~AU \cite{2008ASPC..398..179E}. If systems with Hot Jupiters do host stellar or substellar mass companions, they should lie far enough away from the host star to be detectable. In this way, a survey such as the one we are pursuing will determine whether stellar companions are the cause of the observed spin-orbit misalignment of Hot Jupiter systems and in turn the validity of the Kozai mechanism. 

The contrasts required to detect companions with masses as low as mid M-dwarf and with separations as small as $0.5^{\texttt{"}}$ around solar type stars (i.e. G stars) is $>10^{2}$ in the Ks-band ($\sim2000-2400$~nm). Even higher contrasts ($> 10^{3}$) will be necessary to detect fainter companions (such as late M-dwarf or early L-dwarfs) around solar type stars and to detect companions around more massive and brighter F-type stars. These requirements are readily achievable with current technology. Therefore our group is pursuing a survey of nearby stars ($d \leq 250$~pc) with measured spin-orbit alignments in the southern hemisphere using the Magellan Adaptive Optics (MagAO) and Clio2 infrared camera instruments on the $6.5$~m Magellan Telescope at the Las Campanas Observatory in Chile. The MagAO system can achieve contrasts of $\sim 10^{4}$ ($\Delta Ks = 10$) at $0.5^{\texttt{"}}$ and $\sim 10^{5}$ ($\Delta Ks = 12.5$) at $1.0^{\texttt{"}}$ (private communication L. Close) in the Ks-band (better than other similar surveys \cite{2010ASSL..366...19E}, \cite{2013MNRAS.433.2097F}, \cite{2013MNRAS.428..182B}), that will conclusively confirm or reject the presence of stellar mass companions at separations greater than $150$~AU in our sample.

Other groups are also searching for stellar companions around nearby stars. One such group is the VLT/NACO Search for Stellar Companions to 130 Nearby Stars with and Without Planets survey \cite{2010ASSL..366...19E}. They are using the Nasmyth Adaptive Optics System (NAOS) with the Near-Infrared Imager and Spectrograph (CONICA) instruments on the $8.2$~m Very Large Telescope (VLT) in Cerro Paranal, Chile to search for companions around southern hemisphere stars and the PUEO adaptive optics imager on the $3.6$~m Canada-France-Hawaii Telescope (CFHT) on top of Mauna Kea, Hawaii to search for companions around northern hemisphere stars. Their survey is capable of detecting stellar companions down to M5 dwarfs at $0.2^{\texttt{"}}$ and early L-dwarfs (brown dwarfs) at $0.2^{\texttt{"}}$. Other groups are using the Lucky Imaging technique to search for companions in the SDSS \textit{$i^{\texttt{'}}$} band with the LuckyCam camera on the $2.56$~m Nordic Optical Telescope \cite{2013MNRAS.433.2097F} and in the \textit{$z^{\texttt{'}}$} band with the AstraLux Norte imaging instrument on the Calar Alto $2.2$~m telescope and the AstraLux Sur imaging instrument on the ESO $3.5$~m New Technology Telescope at La Silla, Chile \cite{2013MNRAS.428..182B}.  

Compared to other surveys, ours is unique in that it is the only one that is specifically targeting systems with measured spin-orbit alignments and is directly testing the Kozai mechanism for producing the observed misaligned Hot Jupiters. We also have a key advantage over other surveys as we will be able to achieve higher contrast that will enable us to detect fainter companions.

\section*{Conclusions}
We have presented results revealing that the transiting Hot Jupiter WASP-79b is in significant spin-orbit misalignment. We find that the projected angle between the spin-axis of the host star and the orbital plane of WASP-79b is $\lambda = -106^{+19}_{-13}$\,$^{\circ}$, making its orbit nearly polar. WASP-79b joins the growing population of Hot Jupiters that have been found to exhibit significant spin-orbit misalignments. Several mechanisms have been proposed for producing Hot Jupiters with high obliquities. These include disk migration \cite{2010MNRAS.401.1505B}, Kozai resonances \cite{2011Natur.473..187N}, secular chaos \cite{2011ApJ...735..109W}, planet-planet scattering \cite{2008ApJ...686..580C}, primordial misalignments of proto-planetary disks \cite{2012Natur.491..418B}, and internal gravity wave modulation of the stellar surface of host stars \cite{2012ApJ...758L...6R}. The jury is still out on which mechanism(s) are responsible for producing spin-orbit misalignments. An expansion of the sample of systems with measured obliquities, in particular multiple planet systems, will allow astronomers to determine the dominant mechanisms that produce such misalignments. Systems suitable for Rossiter-McLaughlin effect follow-up observations will come from globally distributed ground-based transit searches such as HATSouth \cite{2013PASP..125..154B}, the recently announced Kepler space telescope K2 mission \cite{2013arXiv1309.0918B}, and new space-based all-sky transit surveys like the Transiting Exoplanet Survey Satellite (TESS; \cite{2009PASP..121..952D}). Finally, we propose a test for the hypothesis that the majority of misaligned Hot Jupiter systems were produced by Kozai resonant behavior resulting from perturbations by as-yet undiscovered binary companions to the planet host stars.

\end{document}